# Quantum Entangled Supercorrelated States in the Jaynes-Cummings Model

A. K. Rajagopal[1], K. L Jensen[1], and F. W. Cummings[2*]

[1]Naval Research Laboratory, Washington D. C. 50378-5320

[2]University of California, Riverside, (Emeritus) CA 92521

The regions of independent quantum states, maximally classically correlated states, and purely quantum entangled (supercorrelated) states described in a recent formulation of quantum information theory by Cerf and Adami are explored here numerically in the parameter space of the well-known exactly soluable Jaynes-Cummings model for equilibrium and nonequilibrium time-dependent ensembles.

The well-known exact solution of the Jaynes - Cummings (JC) model [1, 2] describing the interaction of a single mode radiation field with a two-level atom is the basis for a vast array of the current experiments on foundations of quantum mechanics involving entangled states [3], new ideas for quantum optics [4], and novel device structures such as micromaser [5], single-atom laser etc. [6]. It is also a generic model of the interaction between two dissimilar quantum systems (Bose field and two-level-system, (TLS)) [7, 8]. Thus this model contains all the subtle features of quantum entanglement. A new quantum information theory for entangled systems of Cerf and Adami [9] (CA) mathematically classified (1) independent quantum states, (2) maximally classically correlated states, and (3) a new classically forbidden and therefore purely quantum entangled states called "supercorrelated" states. This is because they found that the conditional entropies of quantum entangled system can be negative, a feature missed in an earlier theory [10]. CA have subsequently explored many further consequences of this discovery [11].

The purpose of this paper is to employ the formal CA theory to elucidate these features numerically in the parameter space of ensembles constructed with the JC model, thus providing an explicit display of the features that CA discovered. We use the exact eigen-solutions of the JC model for constructing two ensembles: (a) thermal (equilibrium) ensemble at a temperature T, where the JC Hamiltonian may be considered as a generic model (Bose-TLS) system and (b) a time-dependent ensemble of quantum optical interest [10, 12]. Two simple models of the radiation often used in the literature, (i) blackbody-like and (ii) pure coherent source [12, 4] are considered in (b). (For a discussion of ensembles, see [13]). For each of these models, we deduce the individual partial density matrices of the radiation and the atomic fields from the system density matrix. This provides us with explicit expressions to examine numerically the negative conditional entropy and information contained therein, as a function of the model parameters, a feature not considered in [9]. For these ensembles, we find the three regions of states described by CA. Unlike CA, we exhibit them in the parameter space of the model and the two ensembles, the temperature and time-scale of the system.

In the notations of [2], the Hamiltonian of the JC model concerns a two-level atom (TLS) interacting with a single given mode of quantized radiation (boson) field of a given frequency, $\omega$, is described in terms of the usual creation, $\hat{a}^\dagger$, and destruction, $\hat{a}$, operators of the boson field; the TLS is represented by the z-component of the Pauli matrix operator with the energy separation of the two atomic levels $\omega_o$, and their mutual interaction is expressed in the rotating wave approximation:

$$H_{A+R} = \hbar\omega\, \hat{a}^\dagger \hat{a} + \frac{\hbar\omega_o}{2}\hat{\sigma}_z + \hbar\kappa\left(\hat{a}^\dagger \hat{\sigma}_- + \hat{a}\hat{\sigma}_+\right). \tag{1}$$

Here $\kappa$ is the dipole-interaction strength between the radiation and the atom or in the generic model, the boson-TLS interaction. Exact solutions of this interacting system are [1, 2]

$$\begin{aligned}
&H_{A+R}|\varphi(n,s)\rangle = \hbar[\Omega(n,s)]|\varphi(n,s)\rangle, (s=1,2)\\
&H_{A+R}|0,g\rangle = -\hbar[\omega_o/2]|0,g\rangle,\\
&|\varphi(n,1)\rangle = Cos\theta_n|n+1,g\rangle + Sin\theta_n|n,e\rangle,\\
&|\varphi(n,2)\rangle = -Sin\theta_n|n+1,g\rangle + Cos\theta_n|n,e\rangle.
\end{aligned} \tag{2}$$

The entangled states $\{|\varphi(n,s)\rangle\}, (s=1,2; n=0,1,2....)$ and $|0,g\rangle$ are orthonormal and complete. Here s=1, 2 labels the entangled states which in their bare condition are the ground (g) and the excited (e) states, and n, the states of the boson field. Also $\Omega(n,s) = \omega(n+1/2) + (3-2s)\lambda_n$, $\tan\theta_n = \kappa\sqrt{(n+1)}/[(\Delta\omega)/2 + \lambda_n]$, $(\Delta\omega) = (\omega - \omega_o)$, and $\lambda_n = \sqrt{(\Delta\omega/2)^2 + \kappa^2(n+1)}$. The angles $\theta_n$ are measures of the entanglement; $\theta_n = 0$ in the noninteracting case (no entanglement) and $\theta_n = \pi/4$ for all n at resonance, $\Delta\omega = 0$. For the sake of simplicity, the numerical results reported in this paper are for the resonant case. In Eq.(2), the eigenstates in terms of the complete set of states of the boson field and the TLS are also given.

**(a) Equilibrium ensemble**: The density matrix constructed by using the maximum entropy principle subject to given average total energy of the system [2, 13], is thus found to be:

$$\rho_{A+R} = \sum_{n=0}^{\infty}\left[|\varphi(n,1)\rangle w(n,1)\langle\varphi(n,1)| + |\varphi(n,2)\rangle w(n,2)\langle\varphi(n,2)|\right] \quad (3)$$
$$+ |0,g\rangle w(0)\langle 0,g|.$$

Here $w(0) = [\exp(\beta\hbar\omega_o/2)]/Z_{A+R}$, $w(n,s) = [\exp-\beta\hbar\Omega(n,s)]/Z_{A+R}$, and $Z_{A+R}$ is the system partition function. Here $\beta = (k_B T)^{-1}$ where T is the temperature. In terms of the boson and TLS states using Eq.(2), it is

$$\rho_{A+R} = \sum_{n=0}^{\infty}\left[(|n+1\rangle\langle n+1|)(|g\rangle\langle g|)\left(w(n,1)Cos^2\theta_n + w(n,2)Sin^2\theta_n\right)\right]$$
$$+ w(0)(|0\rangle\langle 0|)(|g\rangle\langle g|)$$
$$+ \sum_{n=0}^{\infty}\left[(|n\rangle\langle n|)(|e\rangle\langle e|)\left(w(n,1)Sin^2\theta_n + w(n,2)Cos^2\theta_n\right)\right] \quad (4)$$
$$+ \sum_{n=0}^{\infty}(w(n,1) - w(n,2))Cos\theta_n Sin\theta_n$$
$$\left[(|n+1\rangle\langle n|)(|g\rangle\langle e|) + (|n\rangle\langle n+1|)(|e\rangle\langle g|)\right].$$

The third sum in the above represents the entanglement as well as the "decoherence" features of the interacting system. We obtain the "marginal" density matrix

$$\rho_A = Tr_R \rho_{A+R} = f_{(-)}|g\rangle\langle g| + f_{(+)}|e\rangle\langle e|, \quad (5)$$

Here $f_{(-)} = w(0) + \sum_{m=0}^{\infty}\left(w(m,1)Cos^2\theta_m + w(m,2)Sin^2\theta_m\right)$, and $f_{(+)} = 1 - f_{(-)}$ are the occupation probabilities of the g and e states. Similarly, the "marginal" density matrix of the boson is

$$\rho_R = Tr_A \rho_{A+R} = \sum_{n=0}^{\infty} p_n |n\rangle\langle n|, \tag{6}$$

Here $p_0 = w(0) + w(0,1)Sin^2\theta_0 + w(0,2)Cos^2\theta_0$, and for n=1,2,3,.......,
$p_n = w(n,1)Sin^2\theta_n + w(n,2)Cos^2\theta_n + w(n-1,1)Cos^2\theta_{n-1} + w(n-1,2)Sin^2\theta_{n-1}$.

From Eqs. (2, 5, 6), we compute $S_{A+R} = -Tr(\rho_{A+R} \ln \rho_{A+R})$, $S_A = -Tr_A(\rho_A \ln \rho_A)$, and $S_R = -Tr_R(\rho_R \ln \rho_R)$. Henceforth we focus on the resonant case $\omega = \omega_0$ and the dimensionless variables of this ensemble are $(\beta\hbar\omega)^{-1}$ and $(\kappa/\omega)$.

**(b) Non-equilibrium ensemble**: We consider a unitary time-evolution (Liouville - von Neumann) of an initially prescribed density matrix of the system, $\rho_{A+R}(0)$, [1, 10, 12] given by $\rho_{A+R}(t) = U_{A+R}(t)\rho_{A+R}(0)U^\dagger_{A+R}(t)$ where $U_{A+R}(t) = \exp(-itH_{A+R}/\hbar)$. In terms of the eigenstates given in Eq. (2),

$$\rho_{A+R}(t) = |0,g\rangle\langle 0,g|\rho_{A+R}(0)|0,g\rangle\langle 0,g| +$$
$$|0,g\rangle \sum_{m,s'} \langle 0,g|\rho_{A+R}(0)|\varphi(m,s')\rangle [\exp it(\omega_0/2 + \Omega(m,s'))]\langle\varphi(m,s')| + h.c. +$$
$$\sum_{n,s;m,s'} |\varphi(n,s)\rangle\langle\varphi(n,s)|\rho_{A+R}(0)|\varphi(m,s')\rangle [\exp -it(\Omega(n,s) - \Omega(m,s'))]\langle\varphi(m,s')|.$$

$$\tag{7}$$

For simplicity of presentation, we now employ two special initial density matrices, first considered in [12] and later used in [10] to examine interesting aspects of the entropy of the radiation field. There are two radiation models exhibiting different interesting behaviors [12, 4]. The initial state is specified in the form where the photons come from a single mode of radiation considered in [12] and the atom is in its excited (e) state:

$$\rho_{A+R}(0) = \rho_A(0) \otimes \rho_R(0), \quad \rho_A(0) = |e\rangle\langle e|, \quad \rho_R(0) = \sum_n p(n)|n\rangle\langle n| \tag{8}$$

Model (i): Blackbody-like source
$$p(n) = (\overline{N})^n (1+\overline{N})^{-(1+n)}, \quad \overline{N} = \sum_n np(n) = mean\ number\ of\ photons.$$

Model (ii): A single mode pure coherent source [12] for which the photon number distribution is Poissonian, $p_c(n) = (\overline{N})^n \exp(-\overline{N})/n!$, with $\overline{N}$, and $\rho_A(0)$ are as in model (i) . Note that $p_c(n)$ has a peak at $n = \bar{N}$ whereas $p(n)$ of model (i) is monotonic, but both vanish for large n.

Using Eqs. (7, 8) we obtain

$$\rho_{A+R}(t) = \sum_n \begin{Bmatrix} |\varphi(n,1)\rangle p(n)Sin^2\theta_n \langle\varphi(n,1)| + |\varphi(n,2)\rangle p(n)Cos^2\theta_n \langle\varphi(n,2)| + \\ 2|\varphi(n,1)\rangle\{p(n)Sin\theta_n Cos\theta_n Cos[t(\Omega(n,1)-\Omega(n,2))]\}\langle\varphi(n,2)| \end{Bmatrix} \quad (9)$$

It should be emphasized that the preparation of the initial state as in Eq.(8) is nontrivial both conceptually and experimentally. However the above two models have served as guides and we use them here in the same spirit. In contrast to the equilibrium density matrix Eq. (3), the above expression contains only entanglement effects (dependent on $\kappa$ only). From this the "marginal" density matrices are

$$\rho_A(t) = |g\rangle w_g(t)\langle g| + |e\rangle w_e(t)\langle e|, \quad \rho_R(t) = \sum_n |n\rangle P_n(t)\langle n|, \quad (10)$$

where $w_g(t) = \sum_n p(n) W_n(t), \ w_e(t) = 1 - w_g(t)$ and

$P_n(t) = p(n)(1 - W_n(t)) + p(n-1)W_{n-1}(t),$ with $W_n(t) = (Sin^2 2\theta_n)Sin^2(t\kappa\sqrt{(n+1)}).$

The total system entropy in both of these cases is time-independent and solely determined by the initial state density matrix,

$$S_{A+R}(t) = S_{A+R}(t=0) = -\left[\bar{N}\log\bar{N} - (\bar{N}+1)\log(\bar{N}+1)\right] \text{ (model (i))} \quad (11a)$$

$$S_{A+R}(t) = S_{A+R}(t=0) = -\left[\bar{N}\log\bar{N} - \bar{N} - \sum_n p_c(n)\ln(n!)\right] \text{(model (ii))} \quad (11b)$$

By construction this is just the entropy of the given initial radiation field. The counterparts of Eqs.(5, 6) are calculated using Eq. (10) for the two models. For the resonant case the dimensionless parameters for these two ensembles are $\left(\kappa t \ \pi\sqrt{\bar{N}}\right)$ and $\bar{N}$.

The quantum conditional entropies express the residual information in the atomic and radiation systems respectively while retaining the quantum phase information are

$$S(A+R|R) = S_{A+R} - S_R, \quad S(A+R|A) = S_{A+R} - S_A. \quad (12)$$

As shown formally by CA, for quantum entangled subsystems these can be nonmonotonic and may even be negative unlike their counterparts in classical conditional entropies which are non-negative. And finally the quantum mutual entropy is $S(A:R) = S_A + S_R - S_{A+R}$. As in the classical case, $S(A:R) \geq 0$ but can be $\leq 2\min[S_A, S_R]$. When A and R are classically maximally correlated, the <u>classical</u> upper bound $S(A:R) = \min[S_A, S_R]$ is saturated. The range between the classical and quantum upper bounds corresponds to pure quantum entanglement and is called the state of supercorrelation [9]. This is the new feature found by CA and missed entirely in the earlier works [10]. In the present work, we will numerically explore these features as a function of the parameters of the model and the ensembles chosen, by focusing our attention on $S(A+R|R)/S_A$ for simplicity of presentation. Similar results obtain for $S(A+R|A)/S_R$.

Fig.1 shows $(S_{A+R} - S_R)/S_A$ vs. $(\beta\hbar\omega)^{-1}$ for case (a) of the generic boson-TLS system for three representative values of $(\kappa/\omega)$. For $(\beta\hbar\omega)^{-1} \to 0$ the system approaches its ground state, thus $(S_{A+R} - S_R)/S_A \to 1$ for all $(\kappa/\omega)$, as expected. For $(\kappa/\omega)$=0.5, $(S_{A+R} - S_R)/S_A$ remains positive whereas for $(\kappa/\omega)$=2.5 and 5, negative regions (pure quantum entangled supercorrelated states) appear at finite temperatures. The region between the values 1 and zero of $(S_{A+R} - S_R)/S_A$, represents maximally correlated states.

For $(\kappa/\omega)$ greater than 1, the interaction part of the Hamiltonian in Eq. (1) dominates over the TLS part. The eigenvalues $\Omega(n,2)$ then become negative for increasing values of n, as $(\kappa/\omega)$ increases (for the examples considered here, $\Omega(n,2)$ is negative for $(\kappa/\omega)$ =0.5, n=0; for $(\kappa/\omega)$=2.5, n=0, 1, ...5; and for $(\kappa/\omega)$=5, n=0, 1, ...25) and so at low temperatures, the corresponding weights dominate the density matrix, reminiscent of chaotic behavior of the system as $(\kappa/\omega)$ increases. This is analogous to the result of Furuya et. al. [14], displaying chaotic features in a semiclassical version of the JC model, due to entanglement effects. If this result has to be applicable to the radiation-atom system, such strong dipole-interaction strength can only be achieved if the cavity-size in the quantum optics experiment is of nanometer dimensions and the atoms are cooled to very low temperatures as in [15]!

In Fig.2, we display $(S_{A+R} - S_R)/S_A$ vs. $(\kappa t \ \pi\sqrt{\overline{N}})$ for case (b), model (i) for three values of $\overline{N}$. We observe that for $\overline{N} = 1$ (as in the single atom laser case, [6]) $(S_{A+R} - S_R)/S_A$ is negative indicating the appearance of supercorrelated states for many values of $(\kappa t \ \pi\sqrt{\overline{N}})$ whereas for $\overline{N} = 50$, (as for high mean photon number limit, [10]) these supercorrelation states do not appear except in a very small region for low $\left(\kappa t \ \pi\sqrt{\overline{N}}\right)$ values. Indeed, $(S_{A+R} - S_R)/S_A$ for $\left(\kappa t \ \pi\sqrt{\overline{N}}\right)$ near zero has the form, $\left\{\sum_n [(n+1)p(n)]\ln(p(n)/p(n+1))\right\} \Big/ (\overline{N}+1)\left[\ln(\kappa^2 t^2) - 1 + \ln(\overline{N}+1)\right]$ which goes to $0^-$ (the numerator is positive for the two radiation models considered here). This behavior is not perceptible in the Figures displayed. For $\overline{N}$ less than or equal to 5 the supercorrelated states occur but appear to be spread out; they are more spread out for $\overline{N}$ greater than 5, with a decrease in the number of supercorrelated regions as well as exhibit compression of the oscillations. There is no completely disentangled state as in case (a) by virtue of its construction.

In Fig.3, we similarly display the results for model (ii) of case(b), for the same three values of $\overline{N}$. Here again $\overline{N} = 1$ displays large number of supercorrelated regions whereas they decrease for $\overline{N} = 5$. For $\overline{N} = 50,$ we have a very small negative region of $(S_{A+R} - S_R)/S_A$ near zero time-scale as mentioned above, but there is a new feature - oscillations (revivals) for longer times. Similar revivals were found and discussed in [3, 4, 10]. The positive regions of $(S_{A+R} - S_R)/S_A$ correspond to the maximally classically correlated states while we have no approach to the region of disentangled states, as expected. This is a manifestation of the difference between the two radiation models mentioned earlier.

In conclusion, we have shown here by exploring numerically in the parameter space of the exactly solube JC model that the quantum supercorrelated behavior where the marginal entropy exhibits negative values in both the equilibrium (thermal) and nonequilibrium (time-dependent) ensembles in the corresponding parameter space, clearly demonstrating the entanglement effects contained in the JC model. We believe that by this explicit example, we have elucidated the formal results of CA concerning the existence of supercorrelated states in the quantum entangled systems. As noted here, for the JC model their presence depends on the ensemble considered and the interaction strength between the two subsystems. In view of the recent advances in single-atom quantum optics, especially as we approach nanometric cavity (e.g., quantum wells) sizes, we hope these fully quantum correlated states will be explored experimentally in the near future.

We wish to dedicate this paper to the memory of Professor Edwin T. Jaynes. Thanks are due to Dr. R. W. Rendell for drawing our attention to Ref.[13]. AKR and KLJ thank the Office of Naval Research for partial support of their work.

*Present address: 2365 Virginia St. #4, Berkeley, CA 94709-1338
[1]. E. T. Jaynes and F. W. Cummings, Proc. IEEE., **51,** 89 (1963). See also H. Paul, Ann. Der Phys. **11**, 411 (1963) who also reported an exact solution to the JC model but did not explore in full the ramifications of his solution.
[2]. W. H. Louisell, Quantum Statistical Properties of Radiation, John Wiley & Sons, New York (1990). Pp.323 -328.
[3]. S. Haroche, Phys. Today, **51**, 36 (July 1998).
[4]. B. W. Shore and P.L. Knight, in Physics and Probability -Essays in honor of Edwin T. Jaynes, p. 15, eds. W. T. Grandy, Jr. and P. W. Milonni, Cambridge Univ. Press, Cambridge, UK (1993).
[5]. H. Walther, Ibid. p.33.
[6]. M. S. Feld and Kyungwon An, Scientific American, 57 (July 1998); Kyungwon An, R. R. Dasari, and M. S. Feld, in *Atomic and Quantum Optics: High Precision Measurements*, SPIE Proceedings Series, **2799,** 14 (1996). See also G. Raithel et al., in *Cavity Quantum Electrodynamics*, p.57, Edited by P. R. Berman, Academic Press, New York (1994).
[7]. P. Meystre, in *Physics and Probability -Essays in honor of Edwin T. Jaynes,* p.49, eds. W. T. Grandy, Jr. and P. W. Milonni, Cambridge Univ. Press, Cambridge, UK (1993).
[8]. J. H. Eberly, Ibid.p.63.
[9]. N. J. Cerf and C. Adami, Phys. Rev. Lett. **79,** 5194 (1997).
[10]. S. J. D. Phoenix and P. L. Knight, Phys. Rev. **A44**, 6023 (1991).
[11]. N. J. Cerf and C. Adami, Phys. Rev. **A55**, 3371 (1997); Ibid. **A56**, 1721 and 3470 (1997); and Physica **D120**, 62 (1998).
[12]. F. W. Cummings, Phys. Rev. **140**, A1051 (1965).
[13]. J. R. Klauder and E. C. G. Sudarshan, *Fundamentals of Quantum Optics,* W. A. Benjamin, Inc., New York (1968).
[14]. K. Furuya, M. C. Nemes, and G. Q. Pellegrino, Phys. Rev. Lett. **80,** 5524 (1998).
[15] C. J. Hood, M. S. Chapman, T. W. Lynn, and H. J. Kimble, Phys. Rev. Lett. **80,** 4157 (1998).

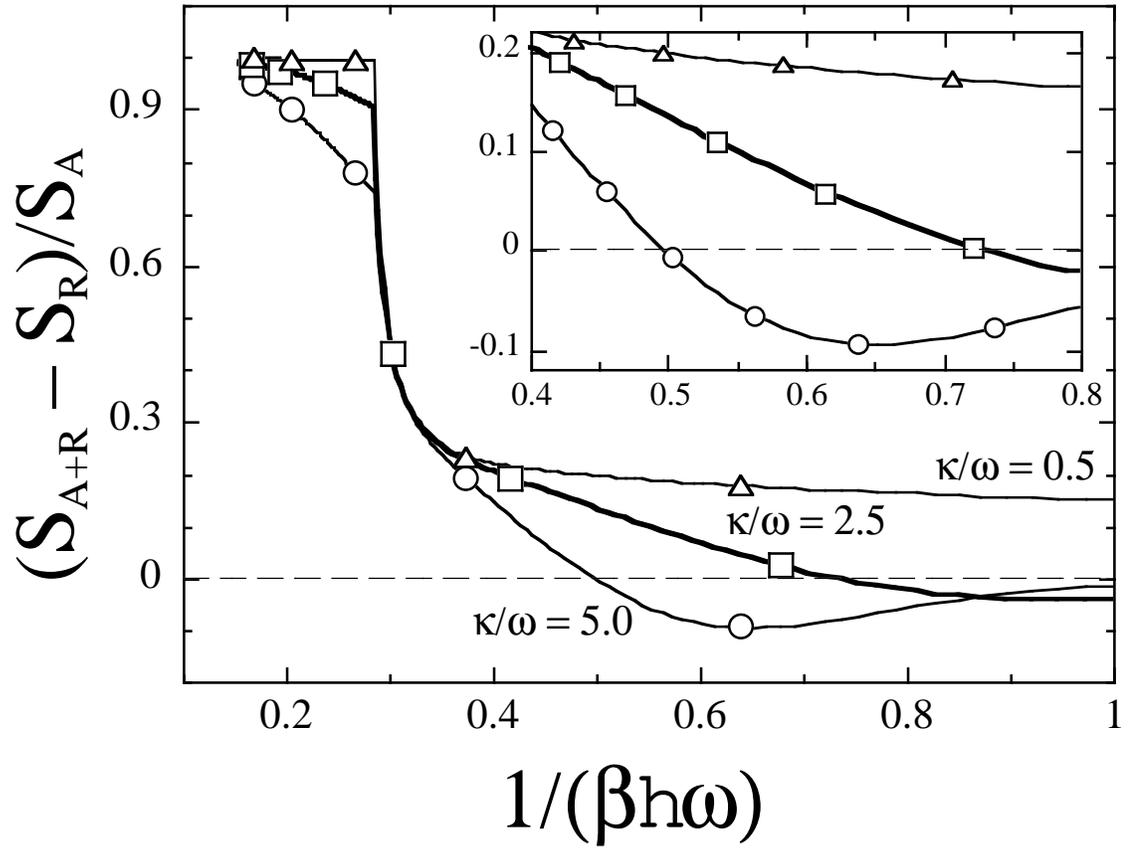

**Figure 1**

Fig. 1: $(S_{A+R} - S_R)/S_A$ vs. $(\beta \hbar \omega)^{-1}$ for three values of the scaled dimensionless coupling parameter, $\kappa/\omega = 0.5, 2.5,$ *and* $5$. The inset exhibits the regions of cross over from positive to negative values of $(S_{A+R} - S_R)/S_A$.

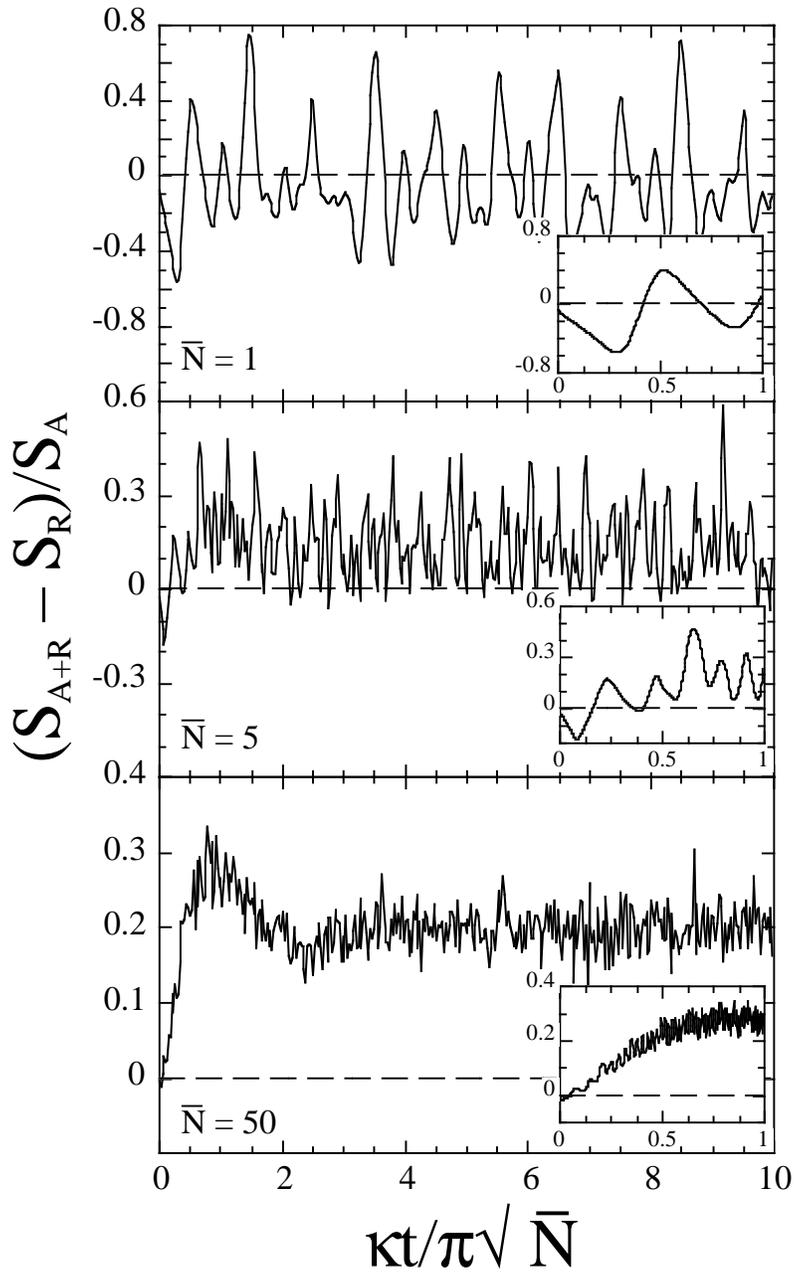

**Figure 2**

Fig. 2: $(S_{A+R} - S_R)/S_A$ vs. $\kappa t / \pi\sqrt{\overline{N}}$ for blackbody-like source for three values of $\overline{N} = 1, 5, and\ 50$. Note that the vertical scales are different for the three different cases. The insets exhibit the cross from negative to positive values of $(S_{A+R} - S_R)/S_A$ for small times.

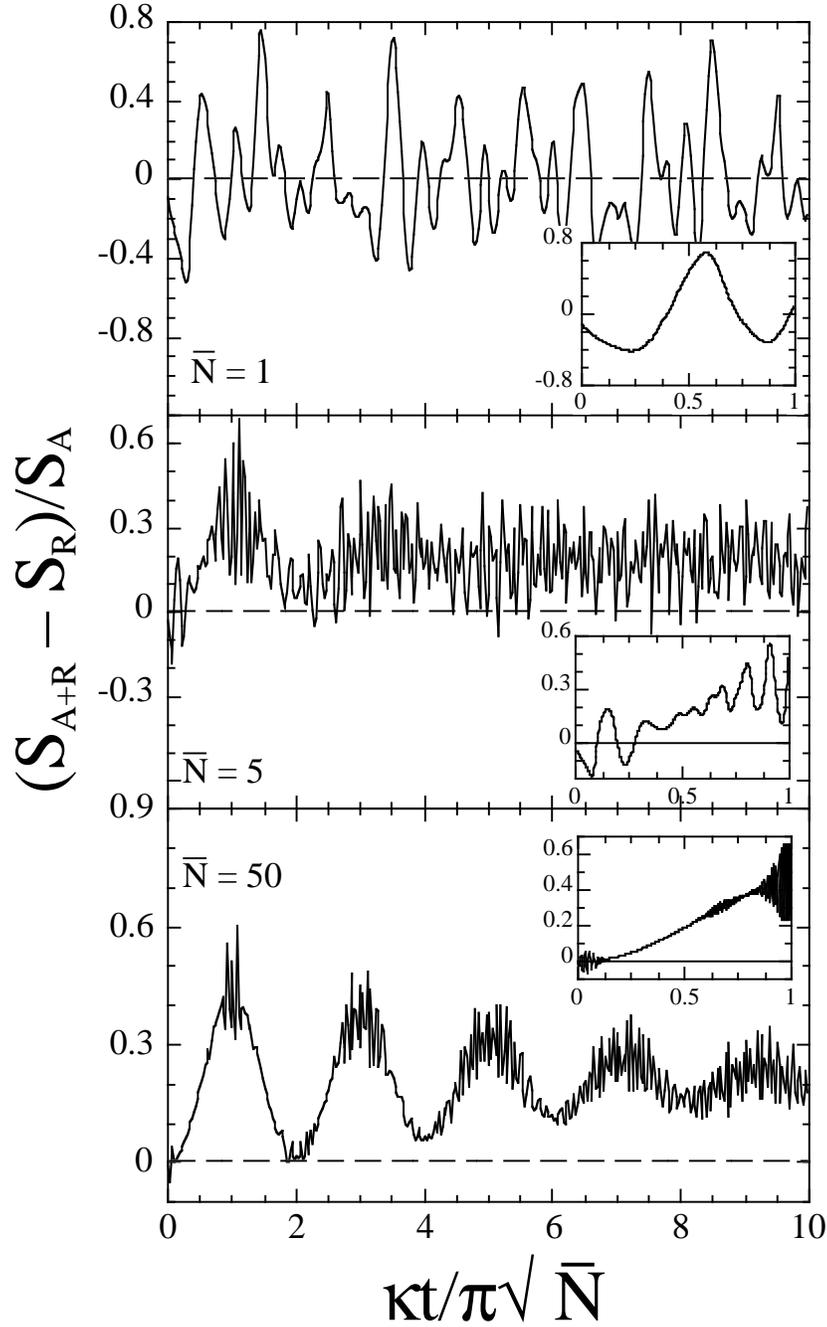

**Figure 3**

Fig. 3: $(S_{A+R} - S_R)/S_A$ vs. $\kappa t \; \pi\sqrt{\overline{N}}$ for pure coherent source for three values of $\overline{N} = 1, 5, and\, 50$. Note that the vertical scales are different for the three different cases. The insets exhibit the cross over from negative to positive values of $(S_{A+R} - S_R)/S_A$ for small times. Unlike in Fig.2, we have oscillations for $\overline{N} = 50$.